\documentclass[aps,prb,twocolumn,showpacs,amsmath,floatfix,superscriptaddress]{revtex4-1}
\usepackage[]{graphicx}
\usepackage{subfigure}
\usepackage{bm}
\usepackage{longtable}
\usepackage{booktabs}
\usepackage{array}
\usepackage[dvipdfm,colorlinks=true,citecolor=blue,linkcolor=blue,urlcolor=blue,bookmarks=true]{hyperref}
\usepackage[sort&compress]{natbib}
\newcommand{\cso}{CuSe$_2$O$_5$ }
\newcommand{\csos}{CuSe$_2$O$_5$}
\newcommand{\fref}[1]{Fig.~\ref{#1}}
\newcommand{\eref}[1]{(\ref{#1})}
\renewcommand{\eqref}[1]{Eq.~(\ref{#1})}
\newcommand{\tref}[1]{Table~\ref{#1}}
\usepackage{color}
\definecolor{red}{rgb}{1,0,0}

\begin{document}
\title{Symmetric and antisymmetric exchange anisotropies in quasi-one-dimensional CuSe$_2$O$_5$ as revealed by ESR}
\author{M. Herak}\email{mirta.herak@ijs.si}
\affiliation{Jo\v{z}ef Stefan Institute, Jamova 39, SI-1000 Ljubljana, Slovenia}
\affiliation{Institute of Physics, Bijeni\v{c}ka c. 46, HR-10000 Zagreb, Croatia}

\author{A. Zorko}
\affiliation{Jo\v{z}ef Stefan Institute, Jamova 39, SI-1000 Ljubljana, Slovenia}
\affiliation{EN-FIST Centre of Excellence, Dunajska 156, SI-1000 Ljubljana, Slovenia}

\author{D. Ar\v{c}on}
\affiliation{Jo\v{z}ef Stefan Institute, Jamova 39, SI-1000 Ljubljana, Slovenia}
\affiliation{Faculty of Mathematics and Physics, University of Ljubljana, Jadranska 19, 1000 Ljubljana, Slovenia}

\author{A. Poto\v{c}nik}
\affiliation{Jo\v{z}ef Stefan Institute, Jamova 39, SI-1000 Ljubljana, Slovenia}

\author{M. Klanj\v{s}ek}
\affiliation{Jo\v{z}ef Stefan Institute, Jamova 39, SI-1000 Ljubljana, Slovenia}
\affiliation{EN-FIST Centre of Excellence, Dunajska 156, SI-1000 Ljubljana, Slovenia}

\author{J. van Tol}
\affiliation{National High Magnetic Field Laboratory, Florida State University, Tallahassee, Florida 32310, USA}

\author{A. Ozarowski}
\affiliation{National High Magnetic Field Laboratory, Florida State University, Tallahassee, Florida 32310, USA}

\author{H. Berger}
\affiliation{Institute of Physics of Complex Matter, EPFL, 1015 Lausanne, Switzerland}

\date{\today}

\begin{abstract}
We present an electron spin resonance (ESR) study of single-crystalline spin chain-system CuSe$_2$O$_5$ in the frequency range between 9~GHz and 430~GHz. In a wide temperature range above the N\'{e}el temperature $T_N=17$~K we observe strong and anisotropic frequency dependence of a resonance linewidth. Although sizeable interchain interaction $J_{IC}\approx 0.1 J$ ($J$ is the intrachain interaction) is present in this system, the ESR results agree well with the Oshikawa-Affleck theory for one-dimensional $S=1/2$ Heisenberg antiferromagnet. This theory is used to extract the anisotropies present in CuSe$_2$O$_5$. We find that the symmetric anisotropic exchange $J_c=(0.04 \pm 0.01) \:J$ and the antisymmetric Dzyaloshinskii-Moriya (DM) interaction $D=(0.05\pm 0.01)\:J$ are very similar in size in this system. Staggered-field susceptibility induced by the presence of the DM interaction is witnessed in the macroscopic susceptibility anisotropy.  
\end{abstract}
\maketitle
\section{Introduction}\label{sec:Intro}
\indent Magnetism of quasi-one-dimensional (1D) $S=1/2$ systems is often well described by an isotropic Heisenberg Hamiltonian. Low dimensionality enhances quantum fluctuations, which in turn suppress long-range ordering. Small interchain interactions present in real systems usually stabilize long-range magnetic ordering at low, but finite, temperatures. The ground state of such systems is, however, very sensitive to the presence of small anisotropy of exchange interactions between spins, frustration and/or defects. Depending on a local symmetry, both symmetric and antisymmetric anisotropic exchange, i.e. the Dzyaloshinskii-Moriya (DM) interaction,\cite{MoriyaPRL-60,*Moriya-60} can be present. In 1D systems where staggered $g$ tensor and/or staggered DM interaction is present, an applied magnetic field induces a staggered field, which opens a gap in the excitation spectrum.\cite{OA-PRL97,OA-PRB99,*OA-PRB00} Determination of the leading anisotropic terms of the spin Hamiltonian thus represents an important milestone in understanding these materials. \\
\indent Magnetic resonance techniques are a very powerful tool for addressing the above points and the type of a ground state in such systems.\cite{KvonNidda-10,Katsumata-00} One of the most appropriate and sensitive methods for determining the presence of small anisotropies is electron spin resonance (ESR),\cite{Zorko-08,Zorko-04} since anisotropic exchange interactions broaden the otherwise exchange narrowed ESR line.\cite{Abragam-Bleaney} Kubo-Tomita (KT) theory is a well-established method of the linewidth analysis, however, it is unfortunately limited only to high temperatures $T\gg J$.\cite{KT-54} Analysis of the ESR linewidth has evolved considerably in the last 15 years, especially regarding the $S=1/2$ 1D Heisenberg antiferromagnet (HAF).\cite{OA-PRL99,OA-02,*OA-07,Choukroun-01} In case of a staggered DM interaction, its contribution to the ESR linewidth is of the same order of magnitude as the symmetric-anisotropy contribution at high temperatures,\cite{Choukroun-01} despite the fact that latter is expected to be smaller, being a higher order perturbation correction to exchange coupling.\cite{MoriyaPRL-60,*Moriya-60} The perturbation theory calculations of the ESR line in $S=1/2$ 1D HAF have been extended to the entire temperature range only for the case of symmetric anisotropic exchange.\cite{Maeda-05} Relatively recently Oshikawa and Affleck (OA) employed field-theory methods to derive a general low-temperature ESR response for half-integer spin 1D HAF.\cite{OA-PRL99,OA-02,*OA-07} Their predictions for the ESR linewidths were experimentally verified in a system with the dominant DM interaction.\cite{Zvyagin-05} Separately, low temperature theories\cite{OA-PRL99,OA-02,*OA-07,Maeda-05} were recently successfully applied to systems with symmetric anisotropic exchange.\cite{Nafradi-10} 
However, quantitative analysis of the low-temperature ESR linewidths for the realistic systems with both anisotropies present has so far been limited.\\
\begin{figure}[tb]
\centering
\includegraphics[clip,width=0.9\columnwidth]{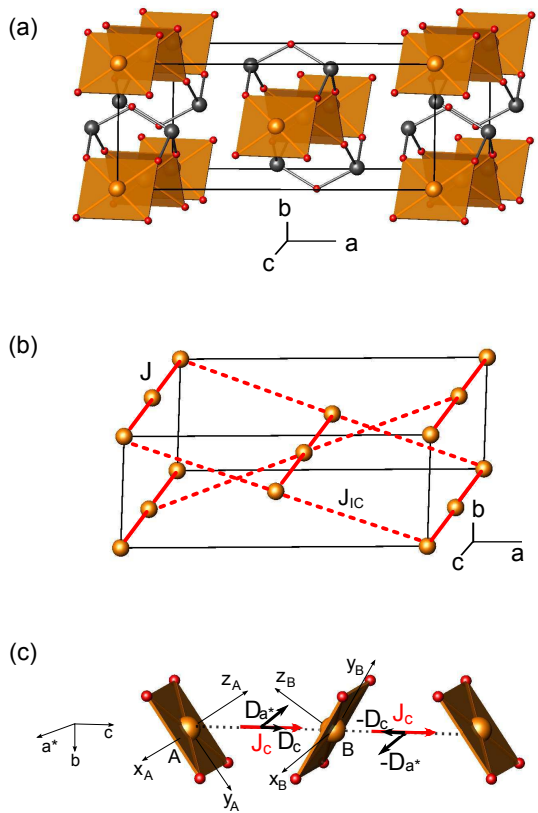}
\caption{(color online). Linear-chain crystal structure in CuSe$_2$O$_5$. (a) Orange plaquettes represent CuO$_4$ rectangles and grey spheres represent Se atoms. (b) Intrachain super-superexchange path with exchange coupling $J$ (solid lines) and dominant interchain path with exchange coupling $J_{IC}$ (dotted lines). (c) Local coordinates of the staggered $g$ tensor, the staggered DM vector $\mathbf{D}=\left(D_{a*},\: 0,\: D_c \right)$ and the axis of the symmetric anisotropic exchange $J_c$. Dotted line represents the chain. }
\label{fig1}
\end{figure}
\indent Among the most studied 1D systems are copper oxides in which magnetism originates from Cu$^{2+}$ spins $S=1/2$ and superexchange is usually mediated through Cu-O-Cu pathways. Copper(II) diselenium(IV) pentoxide is a new 1D copper oxide. It crystallizes in the monoclinic space group $C2/c$.\cite{Meunier-76, Becker-06} The structure consist of $S=1/2$ Cu$^{2+}$ chains running along the crystallographic $c$ axis (\fref{fig1}). Each Se$^{4+}$ ion carries one lone pair of electrons which plays a role of ''chemical scissors''.\cite{Becker-06} The most relevant exchange interactions should be those shown in figures \ref{fig1}(a) and \ref{fig1}(b). The dominant intrachain interaction $J$ is mediated through the double Cu-O-Se-O-Cu super--superexchange (SSE) paths. From a phenomenological (structural) point of view, one may expect significantly smaller interchain (IC) coupling $J_{IC}$ than $J$. The arrangement of the CuO$_4$ plaquettes and the Cu-O-Se-O-Cu bridges is such that it prevents sizeable next-nearest neighbour coupling along the chain. There are two short IC exchange paths, but one, with the modest size of $J_{IC}\approx 0.1\:J$, is expected to be significantly stronger than the other [dotted lines in \fref{fig1}(b)].\cite{Janson-09} \\
\indent The 1D magnetic character of the system is reflected in bulk magnetic susceptibility displaying a broad maximum at $T_{max}=100$~K.\cite{Kahn-80, Janson-09} Above $T_{max}$ it can be well modelled already in the simplest 1D Heisenberg limit, yielding the dominant intrachain exchange $J = 157$~K.\cite{Janson-09} This value is in-line with the Curie-Weiss temperature $\Theta_{CW} = 165$~K and was also successfully theoretically accounted for by density-functional-theory (DFT) calculations, predicting $J = 165$~K and $J_{IC}= 20$~K.\cite{Janson-09} Taking into account the interchain couplings from the DFT calculations, quantum Monte Carlo (QMC) calculations could further improve the agreement between experimental susceptibility and theoretical predictions, although, the disagreement below $T_{max}$ remains noticeable, especially for the chain direction.\cite{Janson-09} Despite being predominantly 1D, the system orders magnetically at $T_N = 17$~K\cite{Janson-09} due to sizable interchain interactions, as evidenced by $J_{IC}\approx T_N$. The susceptibility anisotropy below $T_N$ is consistent with an antiferromagnetic type of spin arrangement, with spins oriented perpendicular to spin chains.\cite{Janson-09} Raman scattering measurements showed that spin-spin correlations emerge below $\approx 110$~K, coinciding with $T_{max}$, and that the system is dominated by enhanced classical spin dynamics as a consequence of a rather strong interchain interaction.\cite{Choi-11} So far \cso was treated as isotropic Heisenberg spin system.\cite{Janson-09,Choi-11} In this work we concentrate on neglected magnetic anisotropies and obtain the anisotropic spin Hamiltonian of the system by simultaneously modelling the angular, the temperature and the frequency dependence of the ESR linewidth with the OA theory. With the obtained anisotropy terms we significantly improve the agreement between measured and modelled susceptibility anisotropy below $T=J$. The results presented here show that both the symmetric and the antisymmetric anisotropic exchange need to be taken into account in case of \csos. In 1D cases when both types of anisotropies are present and comparable, temperature and frequency dependent ESR measurements are invaluable for their quantitative assessment. 
\section{\texorpdfstring{Spin Hamiltonian of C\lowercase{u}S\lowercase{e}$_{\mathbf{2}}$O$_{\mathbf{5}}$}{Spin Hamiltonian in CuSe2O5}}\label{sec:spinHamilt}
\indent The spin Hamiltonian, which describes the spin-spin interactions in a quasi-1D spin system in the applied magnetic field $\mathbf{H}$, is
\begin{equation}\label{eq:hamiltonian}
\mathcal{H} = \mathcal{H}_{iso} + \mathcal{H}_{ae} + \mathcal{H}_{DM}+\mathcal{H}_{Z},
\end{equation}
where
\addtocounter{equation}{-1}
\begin{subequations}
\begin{align}\label{eq:hamiltiso}
	\mathcal{H}_{iso} &=  J\:\sum_{i} \mathbf{S}_i\cdot \mathbf{S}_{i+1} + J_{IC} \: \sum_{<i,j>} \mathbf{S}_i \cdot \mathbf{S}_j,\\\label{eq:hamiltae}
	\mathcal{H}_{ae} &=  J_n \:\sum_{i} S_i^n\:S_{i+1}^n, \\\label{eq:hamiltDM}
	 \mathcal{H}_{DM} &=\sum_{i} \mathbf{D}_i\cdot (\mathbf{S}_i \times \mathbf{S}_{i+1}),\\\label{eq:hamiltZ}
	\mathcal{H}_{Z} &=  -\mu_B\: \sum_{i}\mathbf{S}_{i} \cdot \mathbf{\hat{g}_i}\cdot \mathbf{H}.
\end{align}
\end{subequations}
$\mathcal{H}_{iso}$ is the isotropic Heisenberg interaction with intrachain exchange $J$ and interchain exchange $J_{IC}$. The first sum in $\mathcal{H}_{iso}$ runs over spins along the chain and the second over spins along the chain $\mathbf{S}_i$ and over nearest-neighbouring spins $\mathbf{S}_j$ from two neighbouring chains. $\mathcal{H}_{ae}$ is the symmetric anisotropic exchange with symmetry axis $n$ and relative magnitude $\delta = J_n/J$, $\mathcal{H}_{DM}$ is the antisymmetric DM anisotropic exchange term\cite{MoriyaPRL-60,*Moriya-60} with the site-dependent DM vector $\mathbf{D}_i$ [see \fref{fig1}(c)]. $\mathcal{H}_{Z}$ is the Zeeman term where $\mathbf{\hat{g}}_i$ is the $g$ tensor for site $i$. \\
\indent The orientation of the anisotropic-exchange symmetry axis $n$ may be nontrivial to determine when  two or more inequivalent sites are present.\cite{Eremina-03} In \cso the situation is further complicated by the fact that inequivalent sites A and B [\fref{fig1}(c)] are bridged by the complicated Cu-O-Se-O-Cu SSE path. Nevertheless, below we show that the anisotropic exchange symmetry axis coincides with the chain direction, i.e. $n\equiv c$ in \eqref{eq:hamiltae} [see \fref{fig1}(c)].\\
\indent The general form of the DM vector $\mathbf{D} = (D_{a^*}, 0, D_c)$ is imposed by a two-fold rotational axis along $b$ passing through the middle of each intrachain Cu-Cu bond.\cite{MoriyaPRL-60,*Moriya-60} In addition, the symmetry of \cso is such that the DM interaction is staggered, i.e. $\pm \mathbf{D}$ [\fref{fig1}(c)].\\
\indent The crystal symmetry also dictates the $g$ tensor to be staggered for A and B sites [\fref{fig1}(c)]. For these two sites we thus split $\mathbf{\hat{g}_i}$ in \eqref{eq:hamiltZ} into uniform, $\mathbf{\hat{g}}_u$, and staggered, $\mathbf{\hat{g}}_s$, component and from now on use 
\begin{equation}\label{eq:staggGten}
\mathbf{\hat{g}}_s^{\textup{A,B}}= \mathbf{\hat{g}}_u \pm \mathbf{\hat{g}}_s.	
\end{equation}

\section{Experimental}\label{sec:exp}
\indent Single crystalline \cso samples were synthesized by standard chemical vapour transport method, as described previously, and characterized by X-ray diffraction.\cite{Becker-06}\\
\indent The ESR experiments were performed at X-band (9.4~GHz) and at high-frequencies (HF) between 50~GHz and 430~GHz on single-crystalline samples. Temperature dependence between 4~K and 550~K was measured with temperature stability better than $\pm 0.05$~K. X-band measurements were performed on a home-made spectrometer equipped with the Varian TEM104 dual cavity and the Oxford Instruments ESR900 cryostat.  High-frequency ESR was performed using custom-made transmission type spectrometers at NHMFL facility at Tallahassee, Florida.\cite{HFESR} In all cases the ESR spectra were fitted to a single Lorentzian line. \\
\indent Magnetic susceptibility anisotropy was determined from torque magnetometry measurements performed on a home-made torque apparatus. The resolution of the magnetometer is better than $10^{-4}$~dyn~cm. Measurements were performed in magnetic field of 8~kOe in the temperature range 2 - 330~K.\\
\section{Results}\label{sec:results}
\subsection{X-band ESR}\label{sec:XbandESR}
\begin{figure}[tb]
	\centering
		\includegraphics[width=0.9\columnwidth]{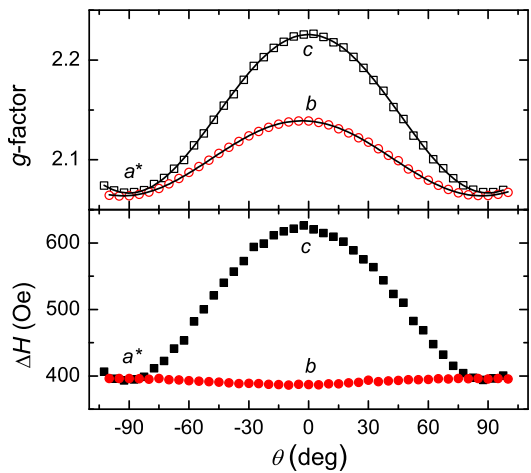}
			\caption{(color online). (upper panel) Angular dependence of the $g$-factor at RT in X-band. (lower panel) Angular dependence of the linewidth at RT in X-band. Solid lines are fits to $g=\sqrt{g_{a^*}^2\sin^2\theta + g_{c,b}^2\cos^2\theta}$.}
				\label{fig2}
\end{figure}

\begin{figure}[t]
	\centering
		\includegraphics[clip,width=0.9\columnwidth]{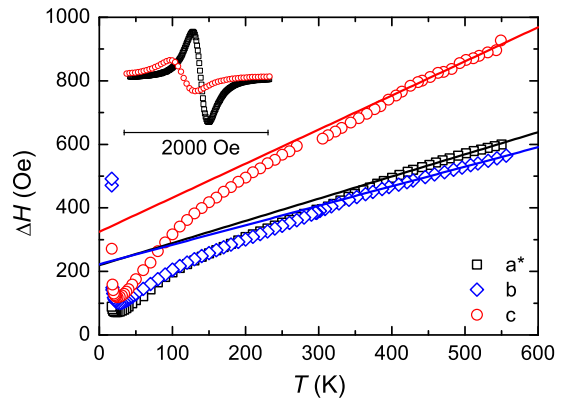}
	\caption{(color online). Temperature dependence of X-band ESR linewidth measured along three crystallographic directions. Solid lines represent linear fits to high temperature data (see text). Inset: RT spectra for $a^*$ and $c$ direction. }
	\label{fig3}
\end{figure}

\indent The room-temperature (RT) angular dependence of the X-band ESR spectra (inset of \fref{fig3}) is shown in \fref{fig2}. It reveals $g$-factor values for the three crystallographically relevant directions, $g_{a^*}=2.064$, $g_{b}=2.140$ and $g_{c}=2.226$. Taking into account the Cu site symmetry we determine the principal eigenvalues of the $g$-tensor  $g_{x}=2.064$, $g_{y}=2.089$ and $g_{z}=2.277$, which are considerably higher than those previously used in theoretical calculations.\cite{Janson-09} The principal axes of the $g$ tensor with respect to the CuO$_4$ plaquette are shown in \fref{fig1}(c). As expected from the crystal structure, local crystal-field symmetry at the copper site is close to being uniaxial with the local anisotropy axis pointing in the direction perpendicular to the CuO$_4$ plaquette, i.e., being tilted by $\alpha=32^{\circ}$ from the $c$ axis around the $a^*$ axis. We take into account that the total measured $g$ tensor is in the strongly exchanged narrowing limit given by $\mathbf{\hat{g}}=(\mathbf{\hat{g}}^{\textup{A}}+ \mathbf{\hat{g}}^{\textup{B}})/2$. The uniform $\mathbf{\hat{g}}_u$ and the staggered component $\mathbf{\hat{g}}_s$ defined in \eqref{eq:staggGten} thus have the following form in the $a^*\:b\:c$ frame
\begin{align}\nonumber
		\mathbf{\hat{g}}_u &=
\begin{pmatrix}
	2.064 & 0 & 0 \\
	0 & 2.140 & 0 \\
	0 & 0 & 2.226
\end{pmatrix}\, ,\\\label{eq:staggG}
\\\nonumber
 \mathbf{\hat{g}}_s &=
\begin{pmatrix}
	0 & 0 & 0 \\
	0 & 0 &  0.084 \\
	0 & 0.084 & 0
\end{pmatrix}\, .
\end{align}
The measured $g$ factors are temperature independent in the paramagnetic state and only slightly increase in the vicinity of $T_N$.\\
\indent At RT, the ESR linewidth anisotropy is pronounced in the $a^*c$ plane and marginal in the plane perpendicular to the chains ($a^*b$ plane). At lower and higher temperatures, however, the anisotropy of $\Delta H$ in the $a^*b$ plane slightly increases, as can be seen in \fref{fig3}. The temperature dependence of the ESR linewidth along crystallographic directions $a^{*}$, $b$ and $c$ was measured from $T_N=17$~K to 550~K (\fref{fig3}). Above $\approx 200$~K the linewidth increases linearly with temperature for all orientations. From the value of $J$ it is not expected that spin-spin correlations would persist up to 550~K, so we attribute this linear dependence to the phonon assisted spin-lattice broadening.\cite{Seehra-68} We fit our high-$T$ data to the phenomenological expression $\Delta H(T)= A + B \: T$ where $\Delta H^{ph}=B\: T$ is the phonon-induced line broadening. Parameter $A$ is the temperature independent exchange-narrowed linewidth, as predicted by the KT theory in the $T\gg J$ regime.\cite{KT-54} The parameters obtained from the fits are summarized in \tref{tab:linTDH}. The values of parameters $A$ and $B$ slightly depend on the temperature range of the fit, which has been accounted for in the parameter errors listed in \tref{tab:linTDH}. \\
\begin{table}[tb]
	\caption{Results of linear fit $\Delta H = A + B \:T$ of high temperature linewidth in $400-550$~K range.}
	\centering
		\begin{tabular}{c||c|c|c}
		\hline
			Direction & $a^*$ & b & c\\ \hline
			$A$~(Oe)  & $220\pm 20$ & $223\pm 7$ & $313\pm 15$\\
			$B$~(Oe/K) & $0.70\pm 0.05$ & $0.61\pm 0.01$ & $1.10\pm 0.06$\\ \hline
		\end{tabular}
	\label{tab:linTDH}
\end{table}
\indent Subtracting the phonon contribution $\Delta H^{ph}$ from the raw data $\Delta H$ results in the corrected linewidth $\Delta H^c=\Delta H - \Delta H^{ph}$, which is determined by the spin-spin interactions only. At high temperatures $\Delta H^c$ is given by the temperature independent coefficients $A$ (\tref{tab:linTDH}) and starts to gradually decrease with decreasing temperature below RT. At $\approx 100$~K there is a crossover to even steeper decrease with decreasing $T$, which is a consequence of the evolution of spin-spin correlations below $T_{max}$.\cite{Janson-09} At $T\approx 25$~K the linewidth has a minimum and then starts to sharply increase with decreasing temperature. Below $T_N$ ESR spectrum disappears at X-band. In principle, the increase of the linewidth in the vicinity of the N\'{e}el point is expected due to critical slowing down of spin fluctuations.\cite{Huber-72} In this case the theory predicts that the linewidth becomes sensitive to the resonance frequency,\cite{Kawasaki-68} which was experimentally seen as a decrease of the linewidth with increasing  frequency.\cite{Seehra-71} This is in contradiction with measurements in \csos, as can be seen in Fig. \ref{fig4}. Here the observed linewidth increases with frequency even at temperatures well above $T_N$, thus suggesting some other origin of field dependent linewidth. OA showed that in 1D $S=1/2$ HAF the linewidth increases with decreasing temperature for $T\ll J$, if staggered fields are present.\cite{OA-PRL99,OA-02,*OA-07} Since in \cso both the staggered $g$ tensor and the staggered DM interaction are potential sources of staggered fields, we now turn to high-frequency ESR results. Nevertheless we note that large increase in X-band linewidths below $\approx 22$~K is reflecting the critical fluctuations in the vicinity of $T_N$.
\subsection{High frequency ESR}\label{sec:HFESR}
\begin{figure}[tb]
	\centering
		\includegraphics[clip,width=0.90\columnwidth]{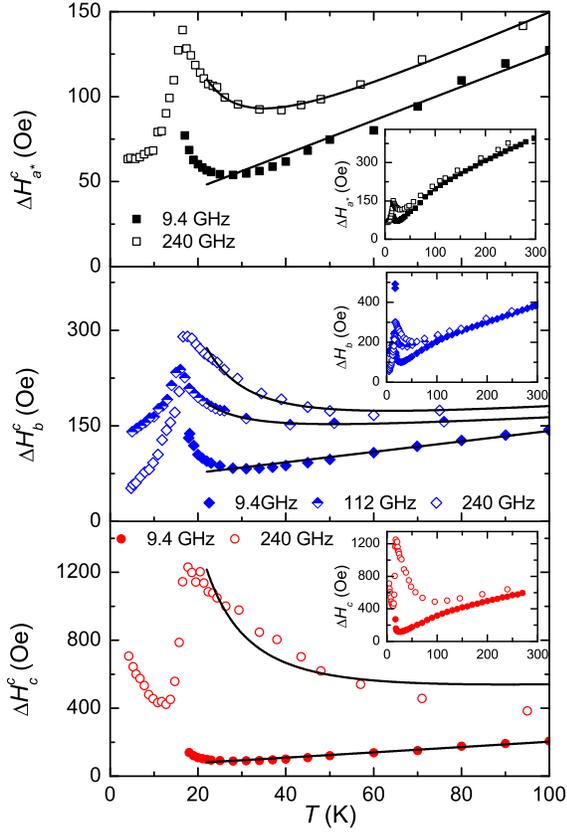}
	\caption{(color online). Temperature dependence of the corrected ESR linewidth $\Delta H^c$ measured at 9.4 GHz, 112 GHz and 240 GHz for magnetic field orientations along all three crystallographic directions. Solid lines are fits to \eqref{eq:totalDH}. Insets: temperature dependence of the raw-data linewidth measured at different frequencies.}
	\label{fig4}
\end{figure}

\indent Temperature dependences of the ESR linewidth measured at 240~GHz for crystal directions $a^*$, $b$ and $c$ are shown and compared to the X-band data in insets of \fref{fig4}. For comparison we also show linewidth measured at 112~GHz for direction $b$. In contrast to the X-band data, at higher frequencies the ESR signal is observable even below $T_N=17$~K, showing a clear anomaly in the linewidth at the transition temperature.\\
\indent It is immediately clear that the magnetic field strongly affects the ESR linewidths along all crystallographic directions. By far the largest effect is seen for the magnetic field along the $c$ direction. To quantitatively analyse the temperature and frequency (field) dependence of the linewidth we first subtracted the high temperature phonon contribution $\Delta H^{ph}$ determined at X-band frequencies. In doing so, we exploit the fact that this contribution is field independent. The resulting linewidths $\Delta H^c$, which will be discussed below, are shown in the main panels of \fref{fig4}.\\
\indent The temperature region where the OA theory is strictly applicable is $T_N\ll T\ll J$, although experiments have been successfully analysed also for $T<J$.\cite{Zvyagin-05} In \cso $J\approx 160$~K and the system orders antiferromagnetically at $T_N= 17$~K, so we limit our analysis to the temperature range $22~\textup{K}<T<100$~K. Increasing the lower limit by a few Kelvins does not influence the extracted parameters. For the Hamiltonian given by \eqref{eq:hamiltonian} the theory for the ESR linewidth in case of symmetric anisotropic exchange predicts \cite{OA-02,*OA-07,Maeda-05}
\begin{equation}\label{eq:OAae}
	\Delta H_{ae}(T) = \dfrac{2\:\epsilon\:k_B\:\delta^2}{g\:\mu_B\:\pi^3}\:T,
\end{equation}
where $\epsilon=2$ applies when magnetic field is along the anisotropy $n$-axis and $\epsilon=1$ otherwise. Low-temperature logarithmic correction can be neglected in the investigated temperature range.\cite{OA-02,*OA-07} \eqref{eq:OAae} predicts linear increase of the ESR linewidth with $T$, which is exactly what we observe for $\Delta H^c$ for all three directions and $T\gg 25$~K (\fref{fig4}). Thus we conclude that the symmetric anisotropic exchange is present in \csos.\\ 
\indent Clear deviations from the linear temperature dependence of $\Delta H^c$ in the low-temperature region imply that the staggered field is also present. Its contribution to the linewidth is given by\cite{OA-PRL99,OA-02,*OA-07}
\begin{equation}\label{eq:OAstaggf}
	\Delta H_{sf}(H,\:T)= 0.69\:g\:\mu_B\:\dfrac{k_B\:J}{(k_B\:T)^2}\:h_s^2\sqrt{\ln\left(\frac{J}{T} \right)}\, ,
\end{equation}
where the staggered field $h_s=c_s\:H$ is proportional to the applied field $H$ and the anisotropic staggered field coefficient $c_s$. \\
\indent Since there are no cross terms between the symmetric and the antisymmetric anisotropic exchange, the total ESR linewidth for 1D system described by Hamiltonian \eref{eq:hamiltonian} is given by
\begin{equation}\label{eq:totalDH}
	\Delta H(H,\:T) = \Delta H_0 + \Delta H_{ae}(T) + \Delta H_{sf}(H,\:T).
\end{equation}
$\Delta H_0$ is the temperature independent linewidth in the high-$T$ limit, which -- according to Refs. [\onlinecite{OA-PRL99,OA-02,*OA-07}] -- contains both the field-independent and the field-dependent contributions. \\
\begin{figure}[tb]
	\centering
		\includegraphics[clip,width=0.9\columnwidth]{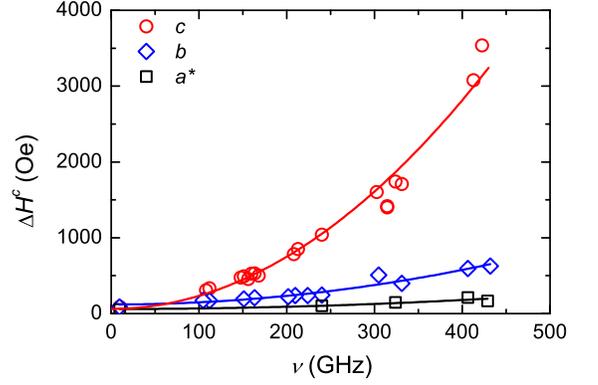}
	\caption{(color online). Frequency dependence of the corrected ESR linewidth $\Delta H^c$ at $T=25$~K. Solid lines are fits to \eqref{eq:freqFit}.}
	\label{fig5}
\end{figure}
\indent Simultaneous fitting of both the temperature (\fref{fig4}) and the frequency dependence (\fref{fig5}) of $\Delta H^c$ to the model given by \eqref{eq:totalDH} allows us to obtain parameters $\Delta H_0$, $\delta$ and staggered field coefficients $c_{s,i}$ for all three magnetic field orientations. The details of the more involved analysis of the frequency dependence are given in Appendix \ref{appA}. Both linewidth dependencies can be fitted (Figs. \ref{fig4} and \ref{fig5}) with the single set of parameters: $c_{s,a^*}=0.009 \pm 0.004$, $c_{s,b}=0.021 \pm 0.004$, $c_{s,c}=0.062 \pm 0.006$, $\delta=0.04\pm0.01$. Simulations also yield that the $c$ axis is the symmetric-anisotropy axis, namely $n=c$ in the Hamiltonian \eref{eq:hamiltae}. The worse agreement is found for X-band data below $\approx 25$~K, which we attribute to the vicinity of the phase transition overshadowing the effect of the induced staggered field. To affirm the correctness of the above fitting parameters we now turn to the low-temperature angular dependence of the linewidth at high frequencies. 
\section{Staggered field analysis and the DM vector}\label{sec:staggField}
\indent In \cso the crystal structure implies that both the staggered $g$ tensor and the staggered DM vector are present, giving the staggered field  
\begin{equation}\label{eq:staggFieldCoeff}
	\mathbf{h}_s = \mathbf{\hat{g}}_u^{-1} \left(\:\mathbf{\hat{g}}_s +\dfrac{1}{2\:J}\:\mathbf{D}\times \mathbf{\hat{g}}_u  \right)\cdot \mathbf{H}.
\end{equation}
The staggered field contribution to the ESR linewidth is determined by the magnitude of $\mathbf{h}_s$, which is proportional to the staggered field coefficients $c_{s}$. For the cases when magnetic field is aligned along the three crystallographic axes, in \cso the staggered field coefficients have the following form 
\begin{align}\nonumber
c_{s,a^*}&=  \dfrac{|D_c| g_{a^*}}{2 g_b} \, ,\\\label{eq:csDM}
c_{s,b} &=\sqrt{\left(\dfrac{D_c g_b}{2 g_{a^*}} \right)^2 + \left( \dfrac{D_{a^*} g_b + 2 g_{bc}}{2 g_c} \right)^2}\, ,\\\nonumber
c_{s,c} &= \left| \dfrac{2 g_{bc} - D_{a^*} g_c}{2 g_b}\right| \, .
\end{align}
The general expression for $c_s$ for arbitrary direction of magnetic field can be found in Appendix \ref{appB}. \\
\begin{figure}[tb]
\includegraphics[clip,width=0.90\columnwidth]{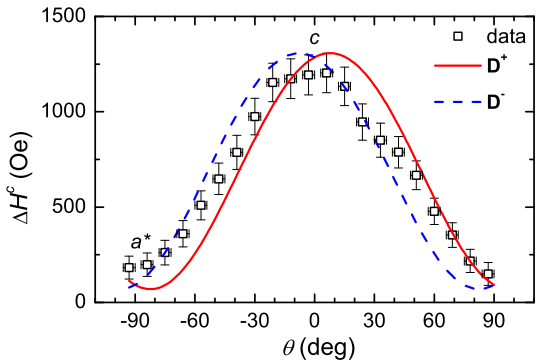}
	\caption{(color online). Angular dependence of ESR linewidth measured at 240~GHz and 25 K in the $a^*c$ plane. Solid and dotted lines are calculated linewidths using \eqref{eq:totalDH} and $\mathbf{D}^+ =(-0.044,\; 0,\; +0.0255)\:J$ and $\mathbf{D}^- =(-0.044,\; 0,\; -0.0255)\:J$, respectively. In both cases $\delta=0.04$. See text for details. }
	\label{fig6}
\end{figure}
\indent Using the experimentally determined $g$ tensor [\eqref{eq:staggG}] and values of $c_{s,i}$, we can now obtain also the DM vector from Eqns. \eref{eq:csDM}. 
Solving $c_{s,a^*}$ and $c_{s,b}$ for $\mathbf{D}$ gives four possible solutions $\mathbf{D}_{1}^{\pm}=(-0.114\pm0.015,\; 0,\; \pm 0.024\pm 0.014)\:J$ and $\mathbf{D}_{2}^{\pm}=(-0.043\pm 0.015,\; 0,\; \pm 0.024\pm 0.014)\:J$. Alternatively, solving $c_{s,c}$ and $c_{s,b}$ gives two real values: $\mathbf{D}_{3}^{\pm}=( -0.045 \pm 0.011,\; 0, \; \pm 0.027\pm 0.014)\:J$. Since DM vectors $\mathbf{D}_{2}^{\pm}$ and $\mathbf{D}_{3}^{\pm}$ nearly coincide, we take this solution as the correct DM vector. 

Finally, we note that the average of $\mathbf{D}_{2}^{\pm}$ and $\mathbf{D}_{3}^{\pm}$, i.e the DM vector $\mathbf{D}=(-0.044 \pm 0.010, \; 0,\; \pm0.0255 \pm 0.010)\:J$, accounts also for the angular dependence of the linewidth measured at 240 GHz and 25 K  (\fref{fig6}). Unfortunately, the present data does not allow us to determine also the sign of the $c$-component of $\mathbf{D}$ as calculated linewidth for both cases describe data equally well.    
\section{Discussion and conclusions}\label{sec:disc}
\begin{figure}[tb]
	\centering
		\includegraphics[width=0.9\columnwidth]{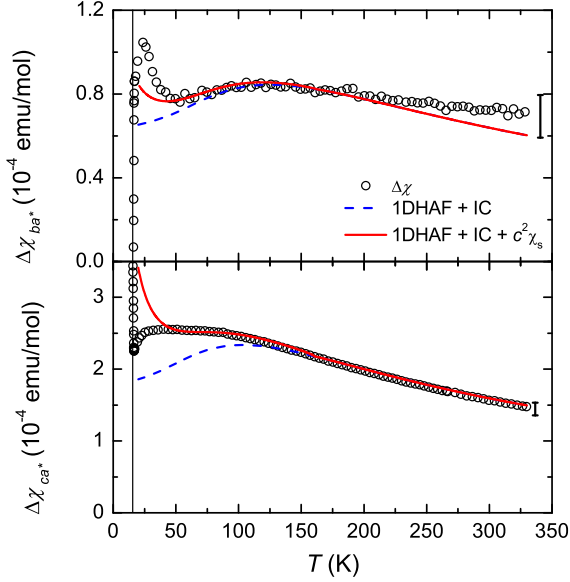}
	\caption{(color online). Susceptibility anisotropy measured in the $a^*b$ plane (upper panel) and $a^*c$ plane (lower panel). Lines represent comparison between susceptibility anisotropy obtained from 1D HAF with interchain interactions (dotted lines) and when staggered susceptibility is included (solid lines). See text for details. The error bar is shown on the right of the data. Vertical line represents the ordering temperature $T_N$.}
	\label{fig7}
\end{figure}

\indent The main experimental finding of this work is the determination of anisotropies in the 1D HAF \csos. Both anisotropies, the symmetric anisotropic exchange and the antisymmetric DM interaction are sizeable and comparable. The former anisotropy is expected to open a gap in the excitation spectrum already in zero magnetic field. This gap is expected to decrease with increasing field.\cite{Kimura-07} However, in contrast at high magnetic fields the staggered field contribution will dominate and define the energy gap. In this limit, the excitation gap is expected to scale with magnetic field, similarly as in other prototypical 1D HAF systems, like Cu-benzoate\cite{Dender-97, OA-PRL97,OA-PRB00} or Cu-pyrimidine dinitrate.\cite{Feyerherm-00,Zvyagin-04} The crossover between the two regimes has to be addressed in the future from both theoretical as well as experimental point of view, since it may lead to unconventional static and dynamic magnetic properties.\\
\indent Presence of both anisotropies should explain the unresolved issue of poor fitting of magnetic susceptibility below $T_{max}$.\cite{Janson-09} Staggered magnetic anisotropy is responsible for the additional anisotropy in the magnetic susceptibility for $T<J$. For this reason we measured the temperature dependence of the magnetic susceptibility anisotropy $\Delta\chi_{ba^*}=\chi_b - \chi_{a^*}$ and $\Delta\chi_{ca^*}=\chi_c - \chi_{a^*}$ by means of torque magnetometry (\fref{fig7}). The measured magnetic susceptibility anisotropies clearly cannot be explained within simple 1D HAF Bonner-Fisher susceptibility\cite{BF-64,Johnston-00} (\fref{fig7}). In calculations we used $J=161$~K, while interchain interactions with $J_{IC}=0.1 J $ and number of nearest neighbouring chains $z=2$ have been taken into account in the mean-field approximation.\cite{Johnston-00}   
The quality of the fit significantly improves when staggered susceptibility $\chi_{s,i}(T)$ is taken into account. $\chi_{s,i}(T)$ is expressed as\cite{OA-PRL97,OA-PRB99,*OA-PRB00}
\begin{equation}\label{eq:staggSusc}
	\chi_{s,i}(T) = 0.2779\:c_{s,i}^2\:\left( \dfrac{N_A\: g^2\mu_B^2}{4\:k_B} \right) \:\dfrac{\sqrt{\ln(J/T)}}{T},
\end{equation}
where $i$ represents the direction of the applied magnetic field. Using above given staggered susceptibility we expand a quantitative agreement with susceptibility anisotropy data down to 50~K with somewhat increased staggered field parameters, $c_{s,a^*}=0.024$, $c_{s,b}=0.047$ and $c_{s,c}=0.13$, the discrepancy between the staggered field parameters obtained from ESR and from susceptibility being similar to what was observed for Cu benzoate \cite{OA-PRB99,*OA-PRB00}. Thus determined magnetic anisotropies even manage to describe the plateau in $\Delta \chi_{ca^*}$ below $T_{max}$ as well as mimic the sudden upturn in $\Delta \chi_{ba^*}$ below 50~K.\\
\indent In conclusion, we studied quasi-one-dimensional \cso with temperature, frequency and angular dependent ESR. ESR linewidth analysis within the OA theory for  $S=1/2$ 1D HAF allowed us to obtain the values of symmetric anisotropic exchange interaction $J_c = (0.04 \pm 0.01)\:J$ and antisymmetric DM interaction $|\mathbf{D}|=(0.05 \pm 0.01) \:J$. \cso thus appears to be an extremely interesting system where both anisotropies are of similar strength, which has some profound effects on the ground state and magnetic susceptibility anisotropies. Present results thus challenge detailed investigations of the low-energy excitation spectrum and the staggered spin susceptibilities by inelastic neutron diffraction or local probe nuclear magnetic resonance techniques. Finally we stress, that this work also demonstrates how to systematically approach ESR data when both anisotropies are present in the 1D system.      
\appendix
\section{\texorpdfstring{Frequency dependence of the linewidth at $T=25$~K}{Frequency dependence of the linewidth at T=25 K}}\label{appA}
\indent Frequency dependence of the linewidth measured at 25~K for all three crystallographic directions is shown in \fref{fig5}. To fit the data to the \eqref{eq:totalDH}, we need to resolve \eqref{eq:totalDH} into frequency-dependent and frequency-independent contributions. The frequency dependent contributions are expected to depend on $\nu^2$, so we fit our data for given direction $i$ to:
\begin{equation}\label{eq:freqFit}
\Delta H _i(T,\nu)	= \alpha_i(T) + \beta_i(T)\: \nu^2, \;\; i=a^*, \; b, \; c,
\end{equation}
where
\addtocounter{equation}{-1}
\begin{subequations}
\begin{align}
\label{eq:alpha}
\alpha_i(T) &= \Delta H_{0,i}(\nu=0) + \Delta H_{ae,\:i}(T), \\
\label{eq:beta}
\beta_i(T)\:\nu^2 &= \Delta H_{0,i}(\nu) +\Delta H_{sf,\:i}(\nu,T).
\end{align}
\end{subequations}
The temperature independent part $\Delta H_0$ in \eqref{eq:totalDH} is also split in two parts, the frequency independent $\Delta H_{0,i}(\nu=0)$ and the frequency dependent $\Delta H_{0,i}(\nu)$. Our intention is to obtain the parameters $\delta$ and $c_s$. However, both parameters $\alpha$ and $\beta$ obtained from the fit consist of two unknown parts. The difficulty arises from the fact that both $\Delta H_{0,i}(\nu=0)$ and $\Delta H_{0,i}(\nu)$ need to be determined first. The origin of these contributions was discussed by OA.\cite{OA-02,*OA-07} From their analysis $\Delta H_{0,i}(\nu=0)$ is expected to be negligibly small because it is proportional to $D^4/J^3$. If we neglect it we obtain $\delta = 0.03$ which is in good agreement with $\delta = 0.04 \pm0.01$ obtained from fits of temperature dependence of the linewidth. To determine the frequency dependent high-$T$ term $\Delta H_{0,i}(\nu)$ we assume that the 9~GHz data is the zero-frequency linewidth limit and subtract it from the 240~GHz linewidth to obtain the frequency-dependent part, $\Delta H_0(\nu=240~\textup{GHz})$. From \eqref{eq:beta} we see that by subtracting this from  $\beta_i(25~\textup{K})\cdot (240$~GHz$)^2$ we obtain $\Delta H_{sf}(\nu=240~\textup{GHz},\: 25~\textup{K})$. Once $\Delta H_{sf}(\nu=240~\textup{GHz},\: 25~\textup{K})$ is obtained the staggered field coefficients can be calculated by using \eqref{eq:OAstaggf}. The results are summarized in \tref{tab:freqFit} and correspond, within errors, to the ones obtained from the fits of the temperature dependence of the linewidth.\\
\begin{table}[tb]
\caption{Results of the fit of the frequency dependence of the linewidth at $T=25$~K shown in \fref{fig5} to \eqref{eq:freqFit}. }
	\footnotesize
	\centering
		\begin{tabular}{c || c | c | c | c }
		\hline
		 & $\alpha$  & $\beta$  & $\Delta H_{0}(240~GHz)$ &  $c_s$\\
		 &  (Oe) &($10^{-3}$~Oe/GHz$^2$) &  (Oe) & \\  \hline
		$a^*$ & $60 \pm 20$ & $0.7 \pm 0.2$ & $20\pm 5$ &  $0.010\pm 0.002$\\
		$b$ & $120 \pm 18$ & $2.9\pm 0.2$ & $20 \pm 5$ &  $0.026\pm 0.001$ \\
		$c$ & $62 \pm 61$ & $17.3 \pm 0.8$ & $90 \pm 5$ &  $0.068 \pm 0.002$\\ 
    \hline		
		\end{tabular}
	\label{tab:freqFit}
\end{table}
\section{\texorpdfstring{Angular dependence of the linewidth at $T=25$~K and $\nu = 240$~GHz}{Angular dependence of the linewidth at T=25 K and f = 240 GHz}}\label{appB}
 Assuming the most general staggered DM vector allowed by symmetry, $\mathbf{D}=(D_{a^*}, 0, D_c)\:J$, the following expression is obtained for the staggered field coefficient $c_s$ from \eqref{eq:staggFieldCoeff}
\begin{align}\label{eq:cs}\nonumber
	c_s^2(\theta, \varphi) &= \left(\dfrac{D_c g_{b} \sin\theta \sin \varphi}{2g_{a^*}}\right)^2 +\\\nonumber
	& +\left(\dfrac{(2g_{bc} - D_{a^*} g_{c})\cos\theta + D_c g_{a^*} \cos \varphi \sin\theta}{2\:g_{b}} \right)^2  + \\
	&+ \left( \dfrac{(D_{a^*} g_{b} + 2g_{bc}) \sin\theta \sin \varphi}{2 g_{c}}\right)^2.
\end{align}
$g$ tensor components can be directly read from \eqref{eq:staggG}, $\theta$ is the polar and $\varphi$ the azimuthal angle (we take $x=a^*$, $y=b$ and $z=c$). \eqref{eq:cs} gives the coefficient $c_s$ for some general direction of the applied field. Experimentally determined staggered field coefficients (Sec. \ref{sec:HFESR}) are equated with \eref{eq:cs} in the following way: $c_{s,a^*}= c_s(\theta=\pi/2, \varphi=0)$, $c_{s,b} = c_s(\theta=\pi/2, \varphi=\pi/2)$ and $c_{s,c} = c_s(\theta=0)$, which gives result \eref{eq:csDM} quoted in the main text.\\ 
\indent Measured angular dependence of the linewidth shown in \fref{fig6} can be compared to \eqref{eq:totalDH} without any free parameters using expression \eref{eq:cs} for the staggered field parameter by setting $\varphi=0$. We assume the following angular dependence of parameter $\epsilon$ in \eref{eq:OAae}: $\epsilon = 1 + \cos^2 \theta$ which gives $\epsilon=1$ for $a^*$ direction and $\epsilon=2$ for $c$, in accord with the previously obtained results. Temperature independent correction $\Delta H_0$ in \eqref{eq:totalDH} is also angular dependent. We simulate this dependence in the following way: $\Delta H_0(\theta)=\Delta H_{0,a^*} + (\Delta H_{0,c}-\Delta H_{0,a^*})\: \cos^2 \theta$, where $\Delta H_{0,a^*}=40$~Oe and $\Delta H_{0,c}=450$~Oe are the temperature independent corrections obtained from fits of \eqref{eq:totalDH} to the temperature dependence of linewidth along the $a^*$ and the $c$ axis, respectively. We also take $\delta =0.04$ obtained from the same fits, and $\mathbf{D}^\pm =(-0.044,\: 0,\: \pm0.0255)\:J$. 
\begin{acknowledgments}
M. H. acknowledges financial support by the Slovene Human Resources Development and Scholarship fund under grant No. 11013-57/2010-5, Postdoc program of the Croatian Science Foundation (Grant No. O-191-2011), and the resources of the Croatian Ministry of Science, Education and Sports under Grant No. 035-0352843-2846. A. Z. and D. A. acknowledge the financial support of the Slovenian Research Agency (project J1-2118 and BI-US/09-12-040).
\end{acknowledgments}
%

\begin{thebibliography}{33}%
\makeatletter
\providecommand \@ifxundefined [1]{%
 \@ifx{#1\undefined}
}%
\providecommand \@ifnum [1]{%
 \ifnum #1\expandafter \@firstoftwo
 \else \expandafter \@secondoftwo
 \fi
}%
\providecommand \@ifx [1]{%
 \ifx #1\expandafter \@firstoftwo
 \else \expandafter \@secondoftwo
 \fi
}%
\providecommand \natexlab [1]{#1}%
\providecommand \enquote  [1]{``#1''}%
\providecommand \bibnamefont  [1]{#1}%
\providecommand \bibfnamefont [1]{#1}%
\providecommand \citenamefont [1]{#1}%
\providecommand \href@noop [0]{\@secondoftwo}%
\providecommand \href [0]{\begingroup \@sanitize@url \@href}%
\providecommand \@href[1]{\@@startlink{#1}\@@href}%
\providecommand \@@href[1]{\endgroup#1\@@endlink}%
\providecommand \@sanitize@url [0]{\catcode `\\12\catcode `\$12\catcode
  `\&12\catcode `\#12\catcode `\^12\catcode `\_12\catcode `\%12\relax}%
\providecommand \@@startlink[1]{}%
\providecommand \@@endlink[0]{}%
\providecommand \url  [0]{\begingroup\@sanitize@url \@url }%
\providecommand \@url [1]{\endgroup\@href {#1}{\urlprefix }}%
\providecommand \urlprefix  [0]{URL }%
\providecommand \Eprint [0]{\href }%
\providecommand \doibase [0]{http://dx.doi.org/}%
\providecommand \selectlanguage [0]{\@gobble}%
\providecommand \bibinfo  [0]{\@secondoftwo}%
\providecommand \bibfield  [0]{\@secondoftwo}%
\providecommand \translation [1]{[#1]}%
\providecommand \BibitemOpen [0]{}%
\providecommand \bibitemStop [0]{}%
\providecommand \bibitemNoStop [0]{.\EOS\space}%
\providecommand \EOS [0]{\spacefactor3000\relax}%
\providecommand \BibitemShut  [1]{\csname bibitem#1\endcsname}%
\let\auto@bib@innerbib\@empty
\bibitem [{\citenamefont {Moriya}(1960{\natexlab{a}})}]{MoriyaPRL-60}%
  \BibitemOpen
  \bibfield  {author} {\bibinfo {author} {\bibfnamefont {T.}~\bibnamefont
  {Moriya}},\ }\href@noop {} {\bibfield  {journal} {\bibinfo  {journal} {Phys.
  Rev. Lett.}\ }\textbf {\bibinfo {volume} {4}},\ \bibinfo {pages} {228}
  (\bibinfo {year} {1960}{\natexlab{a}})}\BibitemShut {NoStop}%
\bibitem [{\citenamefont {Moriya}(1960{\natexlab{b}})}]{Moriya-60}%
  \BibitemOpen
  \bibfield  {author} {\bibinfo {author} {\bibfnamefont {T.}~\bibnamefont
  {Moriya}},\ }\href@noop {} {\bibfield  {journal} {\bibinfo  {journal} {Phys.
  Rev.}\ }\textbf {\bibinfo {volume} {120}},\ \bibinfo {pages} {91} (\bibinfo
  {year} {1960}{\natexlab{b}})}\BibitemShut {NoStop}%
\bibitem [{\citenamefont {Oshikawa}\ and\ \citenamefont
  {Affleck}(1997)}]{OA-PRL97}%
  \BibitemOpen
  \bibfield  {author} {\bibinfo {author} {\bibfnamefont {M.}~\bibnamefont
  {Oshikawa}}\ and\ \bibinfo {author} {\bibfnamefont {I.}~\bibnamefont
  {Affleck}},\ }\href@noop {} {\bibfield  {journal} {\bibinfo  {journal} {Phys.
  Rev. Lett.}\ }\textbf {\bibinfo {volume} {79}},\ \bibinfo {pages} {2883}
  (\bibinfo {year} {1997})}\BibitemShut {NoStop}%
\bibitem [{\citenamefont {Affleck}\ and\ \citenamefont
  {Oshikawa}(1999)}]{OA-PRB99}%
  \BibitemOpen
  \bibfield  {author} {\bibinfo {author} {\bibfnamefont {I.}~\bibnamefont
  {Affleck}}\ and\ \bibinfo {author} {\bibfnamefont {M.}~\bibnamefont
  {Oshikawa}},\ }\href@noop {} {\bibfield  {journal} {\bibinfo  {journal}
  {Phys. Rev. B}\ }\textbf {\bibinfo {volume} {60}},\ \bibinfo {pages} {1038}
  (\bibinfo {year} {1999})}\BibitemShut {NoStop}%
\bibitem [{\citenamefont {Affleck}\ and\ \citenamefont
  {Oshikawa}(2000)}]{OA-PRB00}%
  \BibitemOpen
  \bibfield  {author} {\bibinfo {author} {\bibfnamefont {I.}~\bibnamefont
  {Affleck}}\ and\ \bibinfo {author} {\bibfnamefont {M.}~\bibnamefont
  {Oshikawa}},\ }\href@noop {} {\bibfield  {journal} {\bibinfo  {journal}
  {Phys. Rev. B}\ }\textbf {\bibinfo {volume} {62}},\ \bibinfo {pages}
  {9200(E)} (\bibinfo {year} {2000})}\BibitemShut {NoStop}%
\bibitem [{\citenamefont {von Nidda}\ \emph {et~al.}(2010)\citenamefont {von
  Nidda}, \citenamefont {B$\ddot{\mathrm{u}}$ttgen},\ and\ \citenamefont
  {Loidl}}]{KvonNidda-10}%
  \BibitemOpen
  \bibfield  {author} {\bibinfo {author} {\bibfnamefont {H.-A.~K.}\
  \bibnamefont {von Nidda}}, \bibinfo {author} {\bibfnamefont {N.}~\bibnamefont
  {B$\ddot{\mathrm{u}}$ttgen}}, \ and\ \bibinfo {author} {\bibfnamefont
  {A.}~\bibnamefont {Loidl}},\ }\href@noop {} {\bibfield  {journal} {\bibinfo
  {journal} {The European Phys. J. - Special Topics}\ }\textbf {\bibinfo
  {volume} {180}},\ \bibinfo {pages} {161} (\bibinfo {year}
  {2010})}\BibitemShut {NoStop}%
\bibitem [{\citenamefont {Katsumata}(2000)}]{Katsumata-00}%
  \BibitemOpen
  \bibfield  {author} {\bibinfo {author} {\bibfnamefont {K.}~\bibnamefont
  {Katsumata}},\ }\href@noop {} {\bibfield  {journal} {\bibinfo  {journal} {J.
  Phys.: Cond. Matt.}\ }\textbf {\bibinfo {volume} {12}},\ \bibinfo {pages}
  {R589} (\bibinfo {year} {2000})}\BibitemShut {NoStop}%
\bibitem [{\citenamefont {Zorko}\ \emph {et~al.}(2008)\citenamefont {Zorko},
  \citenamefont {Nellutla}, \citenamefont {van Tol}, \citenamefont {Brunel},
  \citenamefont {Bert}, \citenamefont {Duc}, \citenamefont {Trombe},
  \citenamefont {de~Vries}, \citenamefont {Harrison},\ and\ \citenamefont
  {Mendels}}]{Zorko-08}%
  \BibitemOpen
  \bibfield  {author} {\bibinfo {author} {\bibfnamefont {A.}~\bibnamefont
  {Zorko}}, \bibinfo {author} {\bibfnamefont {S.}~\bibnamefont {Nellutla}},
  \bibinfo {author} {\bibfnamefont {J.}~\bibnamefont {van Tol}}, \bibinfo
  {author} {\bibfnamefont {L.~C.}\ \bibnamefont {Brunel}}, \bibinfo {author}
  {\bibfnamefont {F.}~\bibnamefont {Bert}}, \bibinfo {author} {\bibfnamefont
  {F.}~\bibnamefont {Duc}}, \bibinfo {author} {\bibfnamefont {J.-C.}\
  \bibnamefont {Trombe}}, \bibinfo {author} {\bibfnamefont {M.~A.}\
  \bibnamefont {de~Vries}}, \bibinfo {author} {\bibfnamefont {A.}~\bibnamefont
  {Harrison}}, \ and\ \bibinfo {author} {\bibfnamefont {P.}~\bibnamefont
  {Mendels}},\ }\href@noop {} {\bibfield  {journal} {\bibinfo  {journal} {Phys.
  Rev. Lett.}\ }\textbf {\bibinfo {volume} {101}},\ \bibinfo {pages} {026405}
  (\bibinfo {year} {2008})}\BibitemShut {NoStop}%
\bibitem [{\citenamefont {Zorko}\ \emph {et~al.}(2004)\citenamefont {Zorko},
  \citenamefont {Ar\v{c}on}, \citenamefont {van Tol}, \citenamefont {Brunel},\
  and\ \citenamefont {Kageyama}}]{Zorko-04}%
  \BibitemOpen
  \bibfield  {author} {\bibinfo {author} {\bibfnamefont {A.}~\bibnamefont
  {Zorko}}, \bibinfo {author} {\bibfnamefont {D.}~\bibnamefont {Ar\v{c}on}},
  \bibinfo {author} {\bibfnamefont {H.}~\bibnamefont {van Tol}}, \bibinfo
  {author} {\bibfnamefont {L.~C.}\ \bibnamefont {Brunel}}, \ and\ \bibinfo
  {author} {\bibfnamefont {H.}~\bibnamefont {Kageyama}},\ }\href@noop {}
  {\bibfield  {journal} {\bibinfo  {journal} {Phys. Rev. B}\ }\textbf {\bibinfo
  {volume} {69}},\ \bibinfo {pages} {174420} (\bibinfo {year}
  {2004})}\BibitemShut {NoStop}%
\bibitem [{\citenamefont {Abragam}\ and\ \citenamefont
  {Bleaney}(1970)}]{Abragam-Bleaney}%
  \BibitemOpen
  \bibfield  {author} {\bibinfo {author} {\bibfnamefont {A.}~\bibnamefont
  {Abragam}}\ and\ \bibinfo {author} {\bibfnamefont {B.}~\bibnamefont
  {Bleaney}},\ }\href@noop {} {\emph {\bibinfo {title} {Electron Paramagnetic
  Resonance of Transition Ions}}}\ (\bibinfo  {publisher} {Oxford University
  Press},\ \bibinfo {year} {1970})\BibitemShut {NoStop}%
\bibitem [{\citenamefont {Kubo}\ and\ \citenamefont {Tomita}(1954)}]{KT-54}%
  \BibitemOpen
  \bibfield  {author} {\bibinfo {author} {\bibfnamefont {R.}~\bibnamefont
  {Kubo}}\ and\ \bibinfo {author} {\bibfnamefont {K.}~\bibnamefont {Tomita}},\
  }\href@noop {} {\bibfield  {journal} {\bibinfo  {journal} {J. Phys. Soc.
  Jpn}\ }\textbf {\bibinfo {volume} {9}},\ \bibinfo {pages} {888} (\bibinfo
  {year} {1954})}\BibitemShut {NoStop}%
\bibitem [{\citenamefont {Oshikawa}\ and\ \citenamefont
  {Affleck}(1999)}]{OA-PRL99}%
  \BibitemOpen
  \bibfield  {author} {\bibinfo {author} {\bibfnamefont {M.}~\bibnamefont
  {Oshikawa}}\ and\ \bibinfo {author} {\bibfnamefont {I.}~\bibnamefont
  {Affleck}},\ }\href@noop {} {\bibfield  {journal} {\bibinfo  {journal} {Phys.
  Rev. Lett.}\ }\textbf {\bibinfo {volume} {82}},\ \bibinfo {pages} {5136}
  (\bibinfo {year} {1999})}\BibitemShut {NoStop}%
\bibitem [{\citenamefont {Oshikawa}\ and\ \citenamefont
  {Affleck}(2002)}]{OA-02}%
  \BibitemOpen
  \bibfield  {author} {\bibinfo {author} {\bibfnamefont {M.}~\bibnamefont
  {Oshikawa}}\ and\ \bibinfo {author} {\bibfnamefont {I.}~\bibnamefont
  {Affleck}},\ }\href@noop {} {\bibfield  {journal} {\bibinfo  {journal} {Phys.
  Rev. B}\ }\textbf {\bibinfo {volume} {65}},\ \bibinfo {pages} {134410}
  (\bibinfo {year} {2002})}\BibitemShut {NoStop}%
\bibitem [{\citenamefont {Oshikawa}\ and\ \citenamefont
  {Affleck}(2007)}]{OA-07}%
  \BibitemOpen
  \bibfield  {author} {\bibinfo {author} {\bibfnamefont {M.}~\bibnamefont
  {Oshikawa}}\ and\ \bibinfo {author} {\bibfnamefont {I.}~\bibnamefont
  {Affleck}},\ }\href@noop {} {\bibfield  {journal} {\bibinfo  {journal} {Phys.
  Rev. B}\ }\textbf {\bibinfo {volume} {76}},\ \bibinfo {pages} {109901(E)}
  (\bibinfo {year} {2007})}\BibitemShut {NoStop}%
\bibitem [{\citenamefont {Choukroun}\ \emph {et~al.}(2001)\citenamefont
  {Choukroun}, \citenamefont {Richard},\ and\ \citenamefont
  {Stepanov}}]{Choukroun-01}%
  \BibitemOpen
  \bibfield  {author} {\bibinfo {author} {\bibfnamefont {J.}~\bibnamefont
  {Choukroun}}, \bibinfo {author} {\bibfnamefont {J.-L.}\ \bibnamefont
  {Richard}}, \ and\ \bibinfo {author} {\bibfnamefont {A.}~\bibnamefont
  {Stepanov}},\ }\href@noop {} {\bibfield  {journal} {\bibinfo  {journal}
  {Phys. Rev. Lett.}\ }\textbf {\bibinfo {volume} {87}},\ \bibinfo {pages}
  {127207} (\bibinfo {year} {2001})}\BibitemShut {NoStop}%
\bibitem [{\citenamefont {Maeda}\ \emph {et~al.}(2005)\citenamefont {Maeda},
  \citenamefont {Sakai},\ and\ \citenamefont {Oshikawa}}]{Maeda-05}%
  \BibitemOpen
  \bibfield  {author} {\bibinfo {author} {\bibfnamefont {Y.}~\bibnamefont
  {Maeda}}, \bibinfo {author} {\bibfnamefont {K.}~\bibnamefont {Sakai}}, \ and\
  \bibinfo {author} {\bibfnamefont {M.}~\bibnamefont {Oshikawa}},\ }\href@noop
  {} {\bibfield  {journal} {\bibinfo  {journal} {Phys. Rev. Lett.}\ }\textbf
  {\bibinfo {volume} {95}},\ \bibinfo {pages} {037602} (\bibinfo {year}
  {2005})}\BibitemShut {NoStop}%
\bibitem [{\citenamefont {Zvyagin}\ \emph {et~al.}(2005)\citenamefont
  {Zvyagin}, \citenamefont {Kolezhuk}, \citenamefont {Krzystek},\ and\
  \citenamefont {Feyerherm}}]{Zvyagin-05}%
  \BibitemOpen
  \bibfield  {author} {\bibinfo {author} {\bibfnamefont {S.~A.}\ \bibnamefont
  {Zvyagin}}, \bibinfo {author} {\bibfnamefont {A.~K.}\ \bibnamefont
  {Kolezhuk}}, \bibinfo {author} {\bibfnamefont {J.}~\bibnamefont {Krzystek}},
  \ and\ \bibinfo {author} {\bibfnamefont {R.}~\bibnamefont {Feyerherm}},\
  }\href@noop {} {\bibfield  {journal} {\bibinfo  {journal} {Phys. Rev. Lett.}\
  }\textbf {\bibinfo {volume} {95}},\ \bibinfo {pages} {017207} (\bibinfo
  {year} {2005})}\BibitemShut {NoStop}%
\bibitem [{\citenamefont {N\'{a}fr\'{a}di}\ \emph {et~al.}(2010)\citenamefont
  {N\'{a}fr\'{a}di}, \citenamefont {Olariu}, \citenamefont {Forr\'{o}},
  \citenamefont {M\'{e}zi\`{e}re}, \citenamefont {Batail},\ and\ \citenamefont
  {J\'{a}nossy}}]{Nafradi-10}%
  \BibitemOpen
  \bibfield  {author} {\bibinfo {author} {\bibfnamefont {B.}~\bibnamefont
  {N\'{a}fr\'{a}di}}, \bibinfo {author} {\bibfnamefont {A.}~\bibnamefont
  {Olariu}}, \bibinfo {author} {\bibfnamefont {L.}~\bibnamefont {Forr\'{o}}},
  \bibinfo {author} {\bibfnamefont {C.}~\bibnamefont {M\'{e}zi\`{e}re}},
  \bibinfo {author} {\bibfnamefont {P.}~\bibnamefont {Batail}}, \ and\ \bibinfo
  {author} {\bibfnamefont {A.}~\bibnamefont {J\'{a}nossy}},\ }\href@noop {}
  {\bibfield  {journal} {\bibinfo  {journal} {Phys. Rev. B}\ }\textbf {\bibinfo
  {volume} {81}},\ \bibinfo {pages} {224438} (\bibinfo {year}
  {2010})}\BibitemShut {NoStop}%
\bibitem [{\citenamefont {Meunier}\ \emph {et~al.}(1976)\citenamefont
  {Meunier}, \citenamefont {Svensson},\ and\ \citenamefont
  {Carpy}}]{Meunier-76}%
  \BibitemOpen
  \bibfield  {author} {\bibinfo {author} {\bibfnamefont {P.~G.}\ \bibnamefont
  {Meunier}}, \bibinfo {author} {\bibfnamefont {C.}~\bibnamefont {Svensson}}, \
  and\ \bibinfo {author} {\bibfnamefont {A.}~\bibnamefont {Carpy}},\
  }\href@noop {} {\bibfield  {journal} {\bibinfo  {journal} {Acta Cryst. B}\
  }\textbf {\bibinfo {volume} {32}},\ \bibinfo {pages} {2664} (\bibinfo {year}
  {1976})}\BibitemShut {NoStop}%
\bibitem [{\citenamefont {Becker}\ and\ \citenamefont
  {Berger}(2006)}]{Becker-06}%
  \BibitemOpen
  \bibfield  {author} {\bibinfo {author} {\bibfnamefont {R.}~\bibnamefont
  {Becker}}\ and\ \bibinfo {author} {\bibfnamefont {H.}~\bibnamefont
  {Berger}},\ }\href@noop {} {\bibfield  {journal} {\bibinfo  {journal} {Acta
  Cryst. E}\ }\textbf {\bibinfo {volume} {62}},\ \bibinfo {pages} {i256}
  (\bibinfo {year} {2006})}\BibitemShut {NoStop}%
\bibitem [{\citenamefont {Janson}\ \emph {et~al.}(2009)\citenamefont {Janson},
  \citenamefont {Schnelle}, \citenamefont {Schmidt}, \citenamefont {Prots},
  \citenamefont {Drechsler}, \citenamefont {Filatov},\ and\ \citenamefont
  {Rosner}}]{Janson-09}%
  \BibitemOpen
  \bibfield  {author} {\bibinfo {author} {\bibfnamefont {O.}~\bibnamefont
  {Janson}}, \bibinfo {author} {\bibfnamefont {W.}~\bibnamefont {Schnelle}},
  \bibinfo {author} {\bibfnamefont {M.}~\bibnamefont {Schmidt}}, \bibinfo
  {author} {\bibfnamefont {Y.}~\bibnamefont {Prots}}, \bibinfo {author}
  {\bibfnamefont {S.-L.}\ \bibnamefont {Drechsler}}, \bibinfo {author}
  {\bibfnamefont {S.~K.}\ \bibnamefont {Filatov}}, \ and\ \bibinfo {author}
  {\bibfnamefont {H.}~\bibnamefont {Rosner}},\ }\href@noop {} {\bibfield
  {journal} {\bibinfo  {journal} {New J. Phys.}\ }\textbf {\bibinfo {volume}
  {11}},\ \bibinfo {pages} {113034} (\bibinfo {year} {2009})}\BibitemShut
  {NoStop}%
\bibitem [{\citenamefont {Kahn}\ \emph {et~al.}(1980)\citenamefont {Kahn},
  \citenamefont {Verdaguer}, \citenamefont {Girerd}, \citenamefont {Galy},\
  and\ \citenamefont {Maury}}]{Kahn-80}%
  \BibitemOpen
  \bibfield  {author} {\bibinfo {author} {\bibfnamefont {O.}~\bibnamefont
  {Kahn}}, \bibinfo {author} {\bibfnamefont {M.}~\bibnamefont {Verdaguer}},
  \bibinfo {author} {\bibfnamefont {J.~J.}\ \bibnamefont {Girerd}}, \bibinfo
  {author} {\bibfnamefont {J.}~\bibnamefont {Galy}}, \ and\ \bibinfo {author}
  {\bibfnamefont {F.}~\bibnamefont {Maury}},\ }\href@noop {} {\bibfield
  {journal} {\bibinfo  {journal} {Solid State Commun.}\ }\textbf {\bibinfo
  {volume} {34}},\ \bibinfo {pages} {971} (\bibinfo {year} {1980})}\BibitemShut
  {NoStop}%
\bibitem [{\citenamefont {Choi}\ \emph {et~al.}(2011)\citenamefont {Choi},
  \citenamefont {Lemmens},\ and\ \citenamefont {Berger}}]{Choi-11}%
  \BibitemOpen
  \bibfield  {author} {\bibinfo {author} {\bibfnamefont {K.-Y.}\ \bibnamefont
  {Choi}}, \bibinfo {author} {\bibfnamefont {P.}~\bibnamefont {Lemmens}}, \
  and\ \bibinfo {author} {\bibfnamefont {H.}~\bibnamefont {Berger}},\
  }\href@noop {} {\bibfield  {journal} {\bibinfo  {journal} {Phys. Rev. B}\
  }\textbf {\bibinfo {volume} {83}},\ \bibinfo {pages} {174413} (\bibinfo
  {year} {2011})}\BibitemShut {NoStop}%
\bibitem [{\citenamefont {Eremina}\ \emph {et~al.}(2003)\citenamefont
  {Eremina}, \citenamefont {Eremin}, \citenamefont {Glazkov}, \citenamefont
  {von Nidda},\ and\ \citenamefont {Loidl}}]{Eremina-03}%
  \BibitemOpen
  \bibfield  {author} {\bibinfo {author} {\bibfnamefont {R.~M.}\ \bibnamefont
  {Eremina}}, \bibinfo {author} {\bibfnamefont {M.~V.}\ \bibnamefont {Eremin}},
  \bibinfo {author} {\bibfnamefont {V.~N.}\ \bibnamefont {Glazkov}}, \bibinfo
  {author} {\bibfnamefont {H.-A.~K.}\ \bibnamefont {von Nidda}}, \ and\
  \bibinfo {author} {\bibfnamefont {A.}~\bibnamefont {Loidl}},\ }\href@noop {}
  {\bibfield  {journal} {\bibinfo  {journal} {Phys. Rev. B}\ }\textbf {\bibinfo
  {volume} {68}},\ \bibinfo {pages} {014417} (\bibinfo {year}
  {2003})}\BibitemShut {NoStop}%
\bibitem [{\citenamefont {Hassan}\ \emph {et~al.}(2000)\citenamefont {Hassan},
  \citenamefont {Pardi}, \citenamefont {Krzystek}, \citenamefont {Sienkiewicz},
  \citenamefont {Goy}, \citenamefont {Rohrer},\ and\ \citenamefont
  {Brunel}}]{HFESR}%
  \BibitemOpen
  \bibfield  {author} {\bibinfo {author} {\bibfnamefont {A.~K.}\ \bibnamefont
  {Hassan}}, \bibinfo {author} {\bibfnamefont {L.~A.}\ \bibnamefont {Pardi}},
  \bibinfo {author} {\bibfnamefont {J.}~\bibnamefont {Krzystek}}, \bibinfo
  {author} {\bibfnamefont {A.}~\bibnamefont {Sienkiewicz}}, \bibinfo {author}
  {\bibfnamefont {P.}~\bibnamefont {Goy}}, \bibinfo {author} {\bibfnamefont
  {M.}~\bibnamefont {Rohrer}}, \ and\ \bibinfo {author} {\bibfnamefont {L.~C.}\
  \bibnamefont {Brunel}},\ }\href@noop {} {\bibfield  {journal} {\bibinfo
  {journal} {J. Magn. Res.}\ }\textbf {\bibinfo {volume} {142}},\ \bibinfo
  {pages} {300} (\bibinfo {year} {2000})}\BibitemShut {NoStop}%
\bibitem [{\citenamefont {Seehra}\ and\ \citenamefont
  {Castner}(1968)}]{Seehra-68}%
  \BibitemOpen
  \bibfield  {author} {\bibinfo {author} {\bibfnamefont {M.~S.}\ \bibnamefont
  {Seehra}}\ and\ \bibinfo {author} {\bibfnamefont {T.~G.}\ \bibnamefont
  {Castner}},\ }\href@noop {} {\bibfield  {journal} {\bibinfo  {journal} {Phys.
  Kondens. Materie}\ }\textbf {\bibinfo {volume} {7}},\ \bibinfo {pages} {185}
  (\bibinfo {year} {1968})}\BibitemShut {NoStop}%
\bibitem [{\citenamefont {Huber}(1972)}]{Huber-72}%
  \BibitemOpen
  \bibfield  {author} {\bibinfo {author} {\bibfnamefont {D.~L.}\ \bibnamefont
  {Huber}},\ }\href@noop {} {\bibfield  {journal} {\bibinfo  {journal} {Phys.
  Rev. B}\ }\textbf {\bibinfo {volume} {6}},\ \bibinfo {pages} {3180} (\bibinfo
  {year} {1972})}\BibitemShut {NoStop}%
\bibitem [{\citenamefont {Kawasaki}(1968)}]{Kawasaki-68}%
  \BibitemOpen
  \bibfield  {author} {\bibinfo {author} {\bibfnamefont {K.}~\bibnamefont
  {Kawasaki}},\ }\href@noop {} {\bibfield  {journal} {\bibinfo  {journal}
  {Progr. Theoret. Phys.}\ }\textbf {\bibinfo {volume} {39}},\ \bibinfo {pages}
  {285} (\bibinfo {year} {1968})}\BibitemShut {NoStop}%
\bibitem [{\citenamefont {Seehra}(1971)}]{Seehra-71}%
  \BibitemOpen
  \bibfield  {author} {\bibinfo {author} {\bibfnamefont {M.~S.}\ \bibnamefont
  {Seehra}},\ }\href@noop {} {\bibfield  {journal} {\bibinfo  {journal} {J.
  Appl. Phys.}\ }\textbf {\bibinfo {volume} {42}},\ \bibinfo {pages} {1290}
  (\bibinfo {year} {1971})}\BibitemShut {NoStop}%
\bibitem [{\citenamefont {Kimura}\ \emph {et~al.}(2007)\citenamefont {Kimura},
  \citenamefont {Yashiro}, \citenamefont {Okunishi}, \citenamefont {Hagiwara},
  \citenamefont {He}, \citenamefont {Kindo}, \citenamefont {Taniyama},\ and\
  \citenamefont {Itoh}}]{Kimura-07}%
  \BibitemOpen
  \bibfield  {author} {\bibinfo {author} {\bibfnamefont {S.}~\bibnamefont
  {Kimura}}, \bibinfo {author} {\bibfnamefont {H.}~\bibnamefont {Yashiro}},
  \bibinfo {author} {\bibfnamefont {K.}~\bibnamefont {Okunishi}}, \bibinfo
  {author} {\bibfnamefont {M.}~\bibnamefont {Hagiwara}}, \bibinfo {author}
  {\bibfnamefont {Z.}~\bibnamefont {He}}, \bibinfo {author} {\bibfnamefont
  {K.}~\bibnamefont {Kindo}}, \bibinfo {author} {\bibfnamefont
  {T.}~\bibnamefont {Taniyama}}, \ and\ \bibinfo {author} {\bibfnamefont
  {M.}~\bibnamefont {Itoh}},\ }\href@noop {} {\bibfield  {journal} {\bibinfo
  {journal} {Phys. Rev. Lett.}\ }\textbf {\bibinfo {volume} {99}},\ \bibinfo
  {pages} {087602} (\bibinfo {year} {2007})}\BibitemShut {NoStop}%
\bibitem [{\citenamefont {Dender}\ \emph {et~al.}(1997)\citenamefont {Dender},
  \citenamefont {Hammar}, \citenamefont {Reich}, \citenamefont {Broholm},\ and\
  \citenamefont {Aeppli}}]{Dender-97}%
  \BibitemOpen
  \bibfield  {author} {\bibinfo {author} {\bibfnamefont {D.~C.}\ \bibnamefont
  {Dender}}, \bibinfo {author} {\bibfnamefont {P.~R.}\ \bibnamefont {Hammar}},
  \bibinfo {author} {\bibfnamefont {D.~H.}\ \bibnamefont {Reich}}, \bibinfo
  {author} {\bibfnamefont {C.}~\bibnamefont {Broholm}}, \ and\ \bibinfo
  {author} {\bibfnamefont {G.}~\bibnamefont {Aeppli}},\ }\href@noop {}
  {\bibfield  {journal} {\bibinfo  {journal} {Phys. Rev. Lett.}\ }\textbf
  {\bibinfo {volume} {79}},\ \bibinfo {pages} {1750} (\bibinfo {year}
  {1997})}\BibitemShut {NoStop}%
\bibitem [{\citenamefont {Feyerherm}\ \emph {et~al.}(2000)\citenamefont {Feyerherm},
  \citenamefont {Abens}, \citenamefont {G$\ddot{\mathrm{u}}$nther}, \citenamefont {Ishida}, \citenamefont {Mei{\ss}ner}, \citenamefont {Meschke}, \citenamefont {Nogami},\ and\ \citenamefont {Steiner}}]{Feyerherm-00}%
  \BibitemOpen
  \bibfield  {author} {\bibinfo {author} {\bibfnamefont {R.}\ \bibnamefont
  {Feyerherm}}, \bibinfo {author} {\bibfnamefont {S.}\ \bibnamefont {Abens}},
  \bibinfo {author} {\bibfnamefont {D.}\ \bibnamefont {G$\ddot{\mathrm{u}}$nther}}, \bibinfo
  {author} {\bibfnamefont {T.}~\bibnamefont {Ishida}}, \bibinfo
  {author} {\bibfnamefont {M.}~\bibnamefont {Mei{\ss}ner}}, \bibinfo
  {author} {\bibfnamefont {M.}~\bibnamefont {Meschke}}, \bibinfo
  {author} {\bibfnamefont {T.}~\bibnamefont {Nogami}}, \ and\ \bibinfo
  {author} {\bibfnamefont {M.}~\bibnamefont {Steiner}},\ }\href@noop {}
  {\bibfield  {journal} {\bibinfo  {journal} {J. Phys.: Cond. Matt.}\ }\textbf
  {\bibinfo {volume} {12}},\ \bibinfo {pages} {8495} (\bibinfo {year}
  {2000})}\BibitemShut {NoStop}%
\bibitem [{\citenamefont {Zvyagin}\ \emph {et~al.}(2004)\citenamefont {Zvyagin},
  \citenamefont {Kolezhuk}, \citenamefont {Krzystek},\ and\ \citenamefont {Feyerherm}}]{Zvyagin-04}%
  \BibitemOpen
  \bibfield  {author} {\bibinfo {author} {\bibfnamefont {S.~A.}\ \bibnamefont
  {Zvyagin}}, \bibinfo {author} {\bibfnamefont {A.~K.}\ \bibnamefont {Kolezhuk}},
  \bibinfo {author} {\bibfnamefont {J.}\ \bibnamefont {Krzystek}},\ and\ \bibinfo
  {author} {\bibfnamefont {R.}~\bibnamefont {Feyerherm}},\ }\href@noop {}
  {\bibfield  {journal} {\bibinfo  {journal} {Phys. Rev. Lett.}\ }\textbf
  {\bibinfo {volume} {93}},\ \bibinfo {pages} {027201} (\bibinfo {year}
  {2004})}\BibitemShut {NoStop}%
\bibitem [{\citenamefont {Bonner}\ and\ \citenamefont {Fischer}(1964)}]{BF-64}%
  \BibitemOpen
  \bibfield  {author} {\bibinfo {author} {\bibfnamefont {J.~C.}\ \bibnamefont
  {Bonner}}\ and\ \bibinfo {author} {\bibfnamefont {M.~E.}\ \bibnamefont
  {Fischer}},\ }\href@noop {} {\bibfield  {journal} {\bibinfo  {journal} {Phys.
  Rev.}\ }\textbf {\bibinfo {volume} {135}},\ \bibinfo {pages} {A640} (\bibinfo
  {year} {1964})}\BibitemShut {NoStop}%
\bibitem [{\citenamefont {Johnston}\ \emph {et~al.}(2000)\citenamefont
  {Johnston}, \citenamefont {Kremer}, \citenamefont {Troyer}, \citenamefont
  {Wang}, \citenamefont {Kl$\ddot{\mathrm{u}}$mper}, \citenamefont {Bud'ko},
  \citenamefont {Panchula},\ and\ \citenamefont {Canfield}}]{Johnston-00}%
  \BibitemOpen
  \bibfield  {author} {\bibinfo {author} {\bibfnamefont {D.~C.}\ \bibnamefont
  {Johnston}}, \bibinfo {author} {\bibfnamefont {R.~K.}\ \bibnamefont
  {Kremer}}, \bibinfo {author} {\bibfnamefont {M.}~\bibnamefont {Troyer}},
  \bibinfo {author} {\bibfnamefont {X.}~\bibnamefont {Wang}}, \bibinfo {author}
  {\bibfnamefont {A.}~\bibnamefont {Kl$\ddot{\mathrm{u}}$mper}}, \bibinfo
  {author} {\bibfnamefont {S.~L.}\ \bibnamefont {Bud'ko}}, \bibinfo {author}
  {\bibfnamefont {A.~F.}\ \bibnamefont {Panchula}}, \ and\ \bibinfo {author}
  {\bibfnamefont {P.~C.}\ \bibnamefont {Canfield}},\ }\href@noop {} {\bibfield
  {journal} {\bibinfo  {journal} {Phys. Rev. B}\ }\textbf {\bibinfo {volume}
  {61}},\ \bibinfo {pages} {9558} (\bibinfo {year} {2000})}\BibitemShut
  {NoStop}%
\end{thebibliography}
%
%

\end{document}